\documentclass{desyproc}

\usepackage{hyperref}
\usepackage{setspace}
\usepackage{breqn}

\begin{document}
\title{Magnetic horizons of ultra-high energy cosmic rays}

\author{{\slshape Rafael Alves Batista, G{\"u}nter Sigl} \\[1ex]
II. Institute for Theoretical Physics, University of Hamburg\\ Luruper Chaussee, 149, 22761 Hamburg, Germany\\}

\contribID{xy}

\confID{8648}  
\desyproc{DESY-PROC-2014-04}
\acronym{PANIC14} 
\doi  

\maketitle

\begin{abstract}
The propagation of ultra-high energy cosmic rays in extragalactic magnetic fields can be diffusive, depending on the strength and properties of the fields. In some cases the propagation time of the particles can be comparable to the age of the universe, causing a suppression in the flux measured on Earth. In this work we use magnetic field distributions from cosmological simulations to assess the existence of a magnetic horizon at energies around 10$^{18}$ eV.

\end{abstract}

\section{Introduction}

During their propagation ultra-high energy cosmic rays (UHECRs) can be deflected by the intervening cosmic magnetic fields, namely the extragalactic and galactic. The extragalactic magnetic field has different strengths in different regions of the universe. For instance, in the center of clusters of galaxies it is $\sim$10 $\mu$G, with coherence length of the order of 10 kpc. The existence of magnetic fields in the voids is still controversial \cite{neronov2010}, but there are some indications that they can be $\sim$10$^{-15}$-10$^{-12}$ G, with typical coherence lengths of the order of 1 Mpc \cite{neronov2010}. 

The propagation of cosmic rays in the extragalactic magnetic fields can be diffusive if the scattering length is much smaller than the distance  from the source to the observer. Depending on the magnetic field strength and diffusion length, a significant fraction of these particles can have trajectory lengths comparable to the Hubble radius. In this case, a suppression in the flux of cosmic rays is expected compared to the case in which magnetic fields are absent, leading to the existence of a magnetic horizon for the propagation of cosmic rays. This effect has been previously studied by many authors, including Mollerach \& Roulet~\cite{mollerach2013}, who developed a parametrization for it, under the assumption of Kolmogorov turbulence. In this work we generalize their result for the case of inhomogeneous extragalactic magnetic fields.

\section{Magnetic suppression}

The diffusive cosmic ray spectrum for an expanding universe can be written as~\cite{berezinsky2006}
\begin{equation}
j(E) = \frac{c}{4\pi}  \int\limits_{0}^{z_{max}} dz \left| \frac{dt}{dz} \right| Q(E_g(E,z),z) \frac{dE_g}{dE} \left( \int\limits_{0}^\infty dB  \frac{1}{N_s} \sum\limits_{i=0}^{N_s} \frac{\exp\left( -\frac{r_g^2}{\lambda^2} \right)}{(4\pi\lambda^2)^{3/2}} p(B) \right),
\label{eq:spectot2}
\end{equation}
where $p(B)$ is the probability distribution of the magnetic field strength $B$, $r_g$ is the comoving distance of the source and $\lambda$ is the so-called Syrovatskii variable, given by: 
\begin{equation}
 \lambda^2(E,z,B)= \int\limits_0^z dz' \left| \frac{dt}{dz'} \right| \frac{1}{a^2(z')} \left[  \frac{cl_c(z)}{3} \left(  a_L \left( \frac{E}{E_c(z,B)} \right)^{\frac{1}{3}} + a_H \left( \frac{E}{E_c(z,B)} \right)^2 \right)   \right],
 \label{eq:syro2}
\end{equation}
with $a = 1/1+z$ being the scale factor of the universe and $l_c(z) = l_{c,0} a(z)$ the coherence length of the field at redshift $z$. The parameters $a_L$ and $a_H$  are, respectively, 0.3 and 4. $E_c$ is the critical energy, defined as the energy for which a particle has a Larmor radius equal to the coherence length of the magnetic field.  The probability distribution functions can be obtained from magnetohydrodynamical (MHD) simulations of the local universe. In this work we considered four different cosmological simulations, namely the ones performed by Miniati \cite{miniati2002}, Dolag {\it et al.}~\cite{dolag2005}, Das {\it et al.}~\cite{das2008}, Donnert {\it et al.}~\cite{donnert2009}. 

If the term in parentheses in equation \ref{eq:spectot2} is equal to 1, then the magnetic field dependence will vanish and the shape of the spectrum will be independent of the modes of propagation. This result is known as the propagation theorem~\cite{aloisio2004}, and states that if the separation between the sources in a uniform distribution is much smaller than the characteristic propagation lengths, the UHECR spectrum will have a universal shape. This spectrum ($j_0$) will be henceforth called universal.

We have not considered the actual time evolution of these cosmological simulations. Instead we assume a magnetic field distribution at $z=0$ and extrapolate it to higher redshifts: $B=B_0(1+z)^{2-m}$, with $m$ designating the evolution parameter. Moreover, we assume a Kolmogorov magnetic field with strengths taken from the simulations. 

The suppression factor $G$ can be written as:
\begin{equation}
	G = \frac{j(E)}{j_0(E)} \approx \exp \left[ - \frac{(aX_s)^\alpha}{x^\alpha + bx^\beta}  \right],
\end{equation}
with $x\equiv E/\langle E_c\rangle$, $\alpha$, $\beta$, $a$ and $b$ the best fit parameters obtained by fitting $j(E)/j_0(E)$ with the function in the right-hand side of the equation. The complete list of best fit parameters for these extragalactic magnetic field models can be found in ref.~\cite{alvesbatista2014}. In this expression $X_s = d_s / \sqrt{R_H l_c}$, where $d_s$ is the source separation and $R_H$ the Hubble radius.

\section{Magnetic horizons}

\begin{figure}[h!]
	\centering
	\includegraphics[width=0.58\columnwidth]{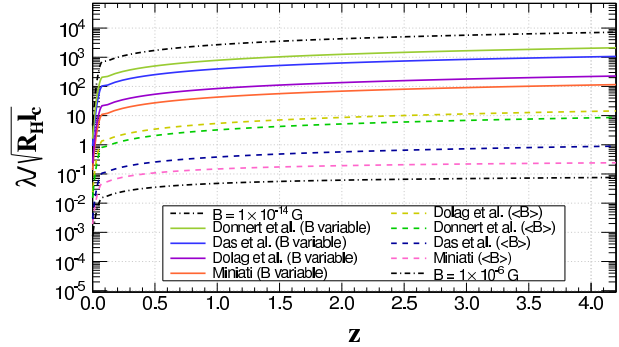}
	\caption{Volume-averaged Syrovatskii variable for an $E/Z=$10$^{16}$ eV, $m=$1, $\gamma=$2 and $z_{max}=$4. Solid lines correspond to the extragalactic magnetic field distribution, dashed lines correspond to the values obtained using the mean magnetic field strengths obtained from  these models, and dotted dashed lines are two limiting cases with high and low magnetic field strengths.}
	\label{fig:maghorizon}
\end{figure}

In this work the magnetic horizon is defined as the mean distance that a cosmic ray can propagate away from the source in a Hubble time. In figure \ref{fig:maghorizon} $\lambda/\sqrt{R_H l_c}$ is displayed as a function of the redshift. In this case  $\lambda$ can be understood as the average distance a particle can propagate away from the source in a time interval corresponding to a redshift $z$.
In this figure we notice that the magnetic horizons for the case of extragalactic magnetic field distributions from cosmological simulations are larger compared to the case of a Kolmogorov turbulent field with $B_{rms}$ equal to the mean magnetic field strength from the distributions. This happens due to the fact that the voids fill most of the volume, dominating the magnetic field distribution and hence the volume-averaged Syrovatskii variable.

\begin{figure}[h!]
	\centering
	\includegraphics[width=0.685\columnwidth]{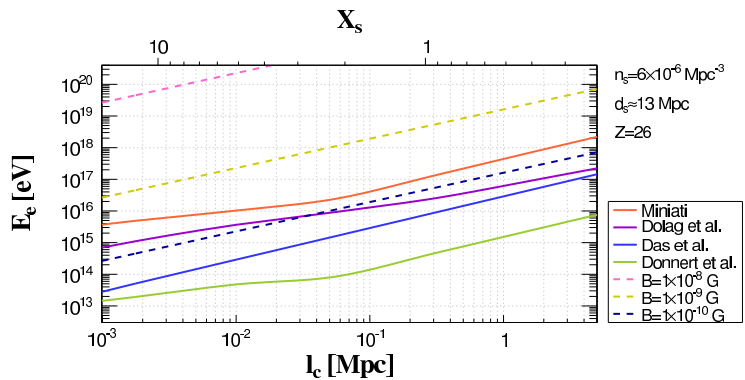}
	\caption{Upper limit on the energy for which the flux of cosmic rays is suppressed to $1/e$ ($\approx$37\%) of its former value, as a function of the coherence length. Solid lines correspond to the indicated extragalactic magnetic field model, and dashed lines to constant magnetic field strengths. This particular case is for a source density of 6$\times$10$^{-6}$ Mpc$^{-3}$ and $Z$=26.}
	\label{fig:upperLimit}
\end{figure}

We can calculate the energy ($E_e$) for which the suppression factor is $G = 1/e \approx 0.37$ of its original value, as a function of the coherence length. The results are shown in figure \ref{fig:upperLimit}.

The magnetic suppression due to magnetic horizon effects starts to become relevant for $E\lesssim$10$^{17}$ eV, for the most optimistic choice of parameters (heavy composition, large coherence length and low source density). The curves in figure \ref{fig:upperLimit} reflect the behavior of the diffusion coefficient, shown in equation \ref{eq:syro2} within square brackets, which is proportional to $l_c^{-1}$  for small values of the coherence length, and to $l_c^{2/3}$ for large $l_c$. 

\section{Discussion and outlook}

We have parametrized the suppression of the cosmic ray flux at energies $\lesssim$Z$\times$10$^{18}$ eV. The method to obtain this parametrization can be adapted to any magnetic field distribution from cosmological simulations (for details see ref.~\cite{alvesbatista2014}). Moreover, we have also derived upper limits for this suppression to occur, as a function of the coherence length.

The results here described suggest that the suppression sets in at energies below $\sim$10$^{17}$ eV. This has profound implications for the interpretation of current experimental data. For instance, recently there has been several attempts \cite{aloisio2013,taylor2014} to perform a combined spectrum-composition fit to data from the Pierre Auger Observatory \cite{auger2010,auger2014}. These results indicate that the spectral indexes of the sources  are hard ($\gamma \sim$1.0-1.6), which contradicts the current acceleration paradigm, in which UHECRs are accelerated to the highest energies through Fermi-like mechanisms ($\gamma \sim$2.0-2.2). In ref.~\cite{mollerach2013} it was shown that the existence of a magnetic horizon around 10$^{18}$ eV can affect the results of these combined fits, softening the spectral index to $\gamma \sim$2. We have shown that if one considers a more realistic extragalactic magnetic field model, the contribution of the voids is dominant and since the field strengths in these regions are low, the suppression will also be small compared to the case of a simple Kolmogorov turbulent magnetic field. In this case, the combined spectrum-composition fits would again favor scenarios in which the sources have hard spectral index.


\section*{Acknowledgements}

RAB acknowledges the support from the Forschungs- und Wissenschaftsstiftung Hamburg.  GS was supported by the State of Hamburg, through the Collaborative Research program ``Connecting Particles with the Cosmos'' and by BMBF under grant 05A11GU1.

\begin{footnotesize}


\end{footnotesize}

\end{document}